# A loss for photoemission is a gain for Auger: Direct experimental evidence of crystal-field splitting and charge transfer in photoelectron spectroscopy

J.C. Woicik and C. Weiland

National Institute of Standards and Technology, Gaithersburg, MD 20899 USA

A.K. Rumaiz

National Synchrotron Light Source, Brookhaven National Laboratory, Upton, NY 11973 USA

We find a new 5 eV satellite in the Ti 1*s* photoelectron spectrum of the transition-metal oxide $SrTiO_3$. This satellite appears in addition to the well-studied 13 eV structure that is typically associated with the Ti 2*p* core line. We give direct *experimental* evidence that the presence of two satellites is due to the crystal-field splitting of the metal 3*d* orbitals. They originate from ligand 2*p* $t_{2g} \rightarrow$ metal 3*d* $t_{2g}$ and ligand 2*p* $e_g \rightarrow$ metal 3*d* $e_g$ monopole charge-transfer excitations within the sudden approximation of quantum mechanics. This assignment is made by the energetics of the resonant and high-energy threshold behaviors of the Ti $K$-$L_2L_3$ Auger decay that follows Ti 1*s* photo-ionization.

The discovery of photoelectron satellites, the structures occurring on the high-binding-energy side of a principle photoelectron-core line, dates back to the early work of Siegbahn [1] and Carlson [2] on rare gases. They are a unique example of the sudden approximation of quantum mechanics, and they illustrate the electron correlations that occur in atoms, molecules, and solids. In fact, the nature and presence of such satellites have been used to establish whether transition-metal compounds are either the charge-transfer or Mott-Hubbard type, being effectively described by both model and Anderson-impurity Hamiltonians [3].

Despite the numerous theoretical descriptions of this unique many-body phenomenon, there has been no direct, experimental evidence that charge is actually transferred (*hops*) from the ligand to the metal ion during the core-photoionization process, a theoretical ansatz that is so central to the physics discussed in these works. For example, it is certain that the satellites occurring in rare gases must be of the shake-up/shake-off origin [4], whereas the chemical bonding in molecules and solids can lead to either "shake-on" or "shake-off."

In this Letter, we employ hard x-ray photoelectron spectroscopy (HAXPES) to study the satellite structure that appears in both the Ti 1*s* and Ti 2*p* core levels of the transition-metal oxide $SrTiO_3$. We find a new photoemission satellite in the Ti 1*s* spectrum that is un-resolved in the Ti 2*p* spectrum due to the spin-orbit splitting of the Ti 2*p* shell. The resonant and high-energy threshold behaviors of the Ti $K$-$L_2L_3$ Auger decay demonstrate that the presence of two satellites is a direct consequence of the crystal-field splitting of the metal 3*d* ion and uniquely identifies them as ligand-to-metal charge transfer. High-energy resonant photoelectron spectroscopy is therefore shown to be a powerful experimental method to study the nature of such transitions because the final state of the resonant decay retains the memory of the initial-state excitation.

Figure 1 shows the Ti 1*s* and Ti 2*p* core-level HAXPES spectra recorded with photon energies $h\nu$ = 5597 eV and $h\nu$ = 4967 eV, respectively, from a $SrTiO_3$ single-crystal surface. Data were recorded at the NIST beamline X24A that is equipped with a Si(111) double-crystal monochromator and a hemispherical electron analyzer. Details of the beamline and vacuum system have been given previously [5]. The sample was etched in buffered-oxide etch for 10 minutes prior to introduction to the vacuum system; it was then annealed at 730º C for 30 minutes.

The well-studied 13 eV Ti satellite [6,7,8,9,10,11] is indicated in both spectra; as seen, it has a binding energy of approximately 13 eV relative to the Ti 1*s*, Ti $2p_{1/2}$, and Ti $2p_{3/2}$ core lines. The Ti 2*p* spectrum is similar to what is reported in the literature, with the exception that our high-resolution spectrum shows a hint of a second low-energy satellite that appears as a shoulder on the high binding-energy side of the $2p_{1/2}$ component and as a shoulder on the low binding-energy side of the 13 eV satellite that is associated with the $2p_{3/2}$ line. The Ti 1*s* spectrum, on the other hand, shows two distinct satellites: The 13 eV structure together with a lower-energy, smaller-intensity feature occurring at approximately 5 eV below the main line. Clearly, the spin-orbit splitting of the Ti 2*p* core level is the reason this lower-energy satellite has not been resolved previously; it is also not resolved in the Ti 2*s* spectrum [8] on account of the anomalously large Coster-Kronig decay width of the 2*s* shell [12].

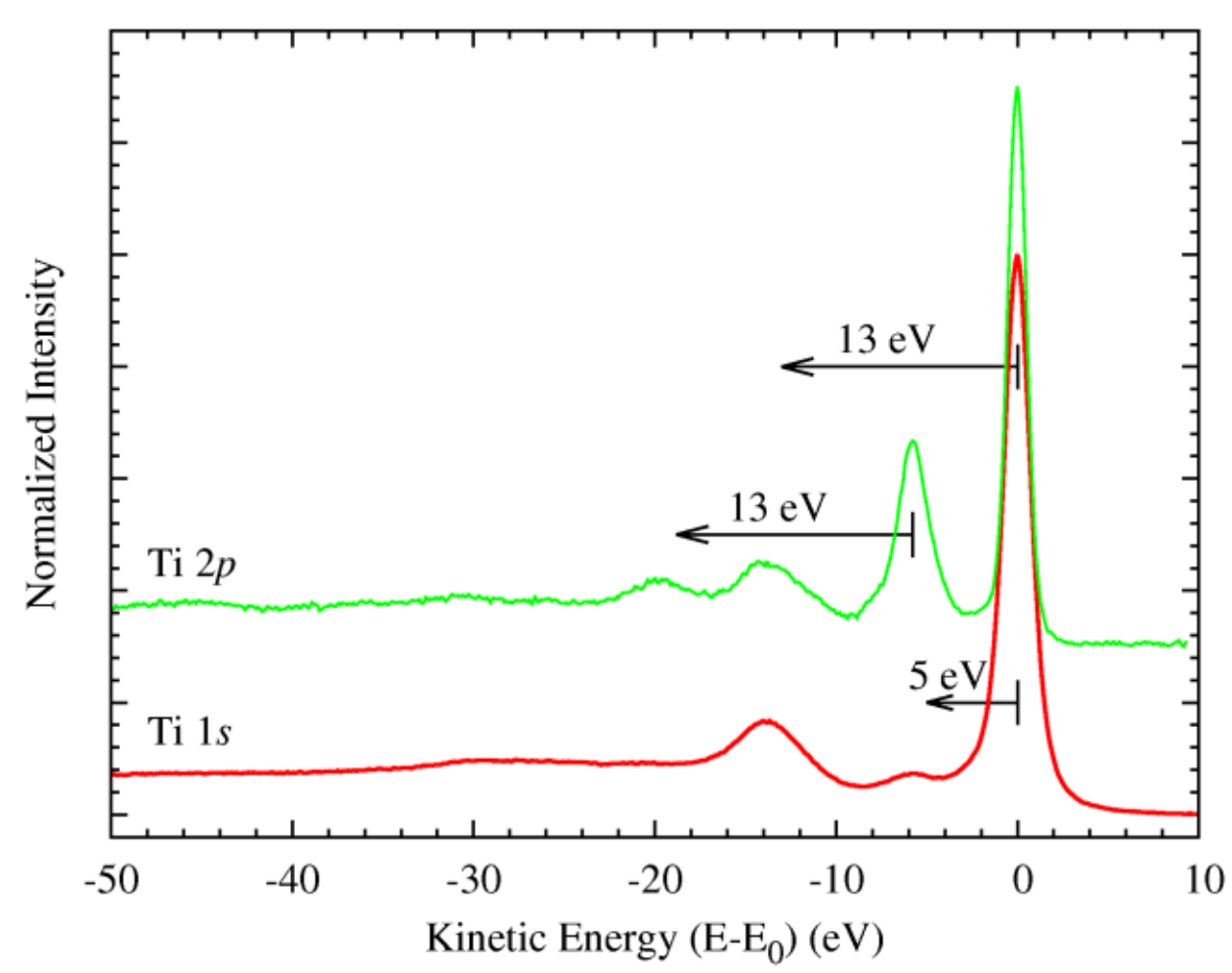

Fig.1. Ti 1*s* photoelectron spectrum recorded with photon energy $h\nu$ = 5597 eV and Ti 2*p* photoelectron spectrum recorded with photon energy $h\nu$ = 4967 eV from cubic $SrTiO_3$. Note the two satellite structures that appear at approximately 5 eV and 13 eV lower-kinetic (higher-binding) energy in the 1*s* spectrum relative to the main 1*s* core line. The higher energy satellite is mirrored in the Ti 2*p* spectrum and occurs at approximately the same energies relative to the Ti $2p_{1/2}$ and the Ti $2p_{3/2}$ core lines. The data have been normalized to equal peak height.

Before we present our resonant data, it is important to elucidate the different electronic transitions that will be studied. Figure 2 shows Ti *K* x-ray-absorption near-edge spectra for single-crystal $SrTiO_3$ [13]. The data are plotted for different sample geometries relative to the incident synchrotron-beam wave vector ***q*** and synchrotron-beam polarization vector ***e***. By orienting the direction of the electric-field polarization and wave vector, the different dipole and quadrupole transitions of the Ti 1*s* electron may be selected as shown in the figure.

In cubic materials, such as $SrTiO_3$, the intensity of dipole transitions is invariant with respect to ***q*** and ***e*** [14]. As shown by their sensitivity to sample geometry, the first two peaks of the spectra are dipole-forbidden transitions of the Ti 1*s* electron to the Ti 3*d* derived $t_{2g}$ ($d_{xy}$, $d_{yz}$, and $d_{zx}$) and $e_g$ ($d_{3z^2-r^2}$, and $d_{x^2-y^2}$) unoccupied molecular orbitals. The energy difference between the two peaks is approximately 2.2 eV, which corresponds to the crystal-field splitting of the metal 3*d* shell. This splitting results from the different orbital overlap between the Ti 3*d* orbitals and the ligand 2*p* orbitals that are strong functions of symmetry; it stabilizes (destabilizes) the occupied (unoccupied) $e_g$ orbitals more strongly than the occupied (unoccupied) $t_{2g}$ orbitals. In a simple ionic picture, this is because the metal $e_g$ orbitals point towards the ligand atoms while the metal $t_{2g}$ orbitals point between them [15].

Note that the $t_{2g}$ and $e_g$ transitions are mirrored at higher-photon energy (by approximately 5 eV), and we have indicated these transitions as $t'_{2g}$ and $e'_g$, respectively, in the figure. These higher-energy transitions, however, show no geometry dependence, indicating that they are dipole allowed. In the theoretical work of Vedrinskii et al. [16] and Vankó et al. [17], similar features in both the $SrTiO_3$ and $LaCoO_3$ spectra were attributed to 1*s* transitions to the metal 3*d* orbitals on neighboring metal atoms via oxygen-mediated intersite hybridization: M(4*p*)–O(2*p*)–M′(3*d*). The fact that they appear at higher-photon energy than the direct (local) 1*s* to 3*d* transitions is attributed to the reduced core-hole attraction on the neighboring metal sites, although their increased energy can also be explained in a simple molecular-orbital picture as the metal 4*p* orbitals lie at higher energy than the metal 3*d* orbitals and hence so would their hybrid. The fact that these transitions are dipole allowed and show such small intensity in the absorption spectra also demonstrates that most of their character originates from the neighboring metal-ion 3*d* states. Note that all of these features lie well below the main 1*s* to 4*p* absorption edge that occurs at 4984 eV in $SrTiO_3$

In our resonant measurements, we set the photon energy to the energies of the above transitions and measure high-resolution Ti *K*-$L_2L_3$ Auger spectra. As in the study of Danger et al. [18], we report only the $^1D_2$ *K*-$L_2L_3$ peak as it is the most intense and other peaks exhibit similar behavior. Figure 3 shows Ti *K*-$L_2L_3$($^1D_2$) Auger spectra recorded with the photon energy set to the 1*s* → $t_{2g}$ quadrupole transition ($h\upsilon$ = 4968.3 eV), the 1*s* → $t'_{2g}$ dipole transition ($h\upsilon$ = 4973.7 eV), and the 1*s* → 4*p* dipole transition ($h\upsilon$ = 4984.0 eV), as indicated. Note the similar kinetic energies of the Auger transition recorded at the latter two photon energies, despite the fact that the former of the two transitions lies below the primary 4*p* edge or "white line" and would therefore typically be considered a bound state. Note as well the large energy shift (by a full 2 eV) of the Auger peak recorded with the photon energy set to the quadrupole 1*s* → $t_{2g}$ transition; this shift is the same when the photon energy is set to the quadrupole 1*s* → $e_g$ transition (not shown), and it is also noticeably narrower for both transitions as expected [19,20].

Clearly, these data demonstrate the localized nature of the metal 3*d* states in this transition-metal oxide. Promotion of the spectator electron to either the metal 3*d* $t_{2g}$ or the metal 3*d* $e_g$ orbitals is both sufficiently localized and long lived to fully screen the photo-hole, and the Auger decay in this case more closely resembles direct photo-ionization because its final state leaves only one photo-hole on the Ti ion. In the case of the two dipole transitions, the strong intersite hybridization delocalizes the spectator electron, leaving little or no spectator-electron density on the absorbing atom with which to screen the photo-hole. It is interesting to note that the energy of the resonant Auger decay recorded at the nonlocal 1*s* → 3*d*′ transition occurs at slightly lower kinetic energy than at the 1*s* → 4*p* transition, again consistent with the assignment that the latter transition is to orbitals that have the majority of their character on neighboring metal sites. Such dispersive energy shifts have been used previously to study electron dynamics in adsorbed systems [21,22,23].

Figure 4 again shows high-resolution Ti *K*-$L_2L_3$ Auger spectra, but now plotted as a function of excess photon energy above the 1*s* → 4*p* threshold as indicated. Note that the energy of the 1*s* → 4*p* threshold is a full 15 eV above the 1*s* → $t_{2g}$ transition, and consequently it is well above the resonant Raman regime that has been studied in the past [20,24,25,18]; our high-energy threshold data therefore probe the electron dynamics that occur as the Ti 1*s* electron transits to the continuum, as opposed to the resonant behavior that occurs when it is trapped in the 3*d* bound state below it [26].

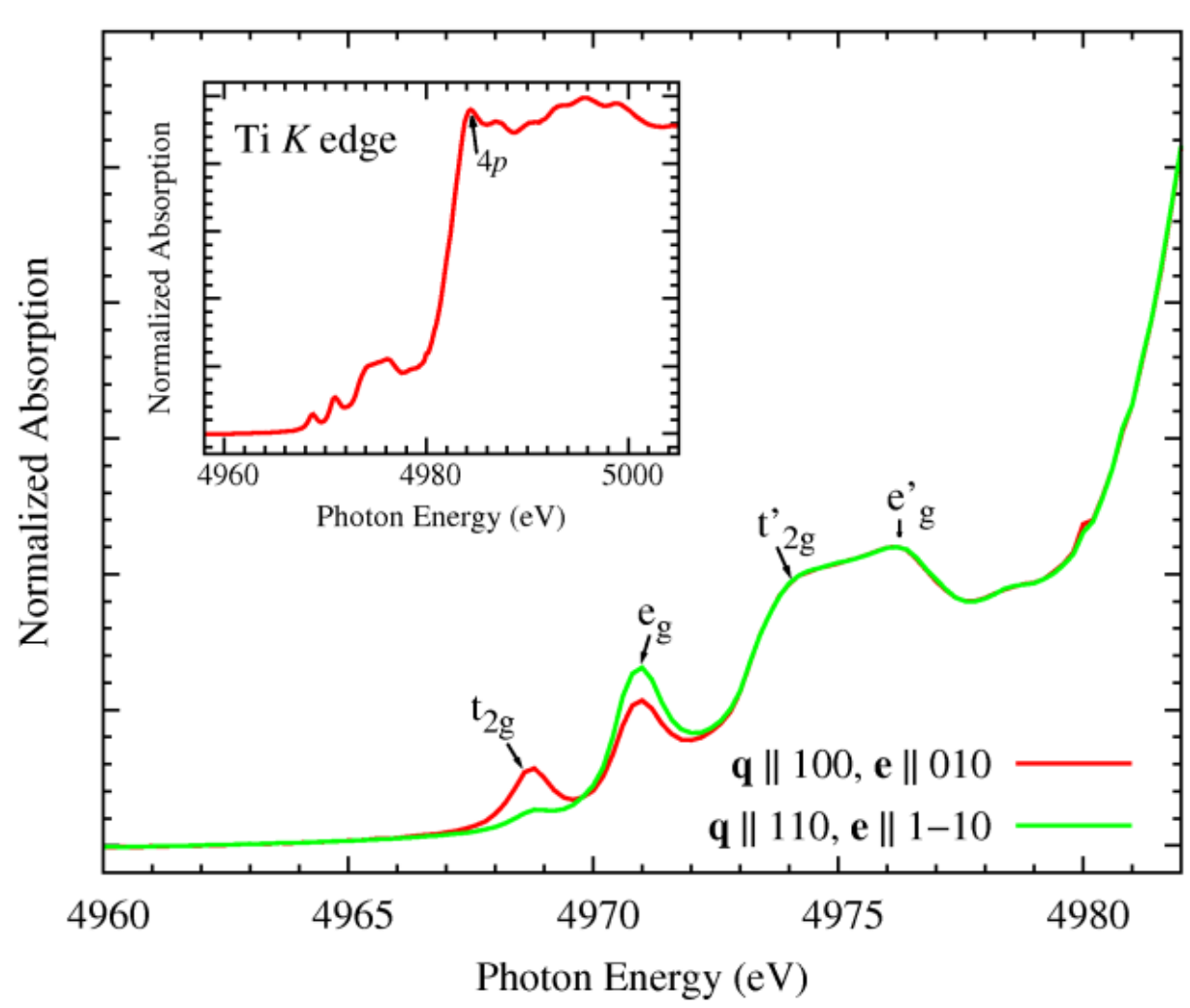

Fig.2. Polarization dependence of the Ti *K* (1*s*) x-ray absorption spectra for cubic $SrTiO_3$ showing the pre-edge structures below the 1*s* → 4*p* threshold. The inset shows the full near-edge region. Transitions to the $t_{2g}$, $e_g$, $t'_{2g}$, $e'_g$, and 4*p* levels are indicated (see text).

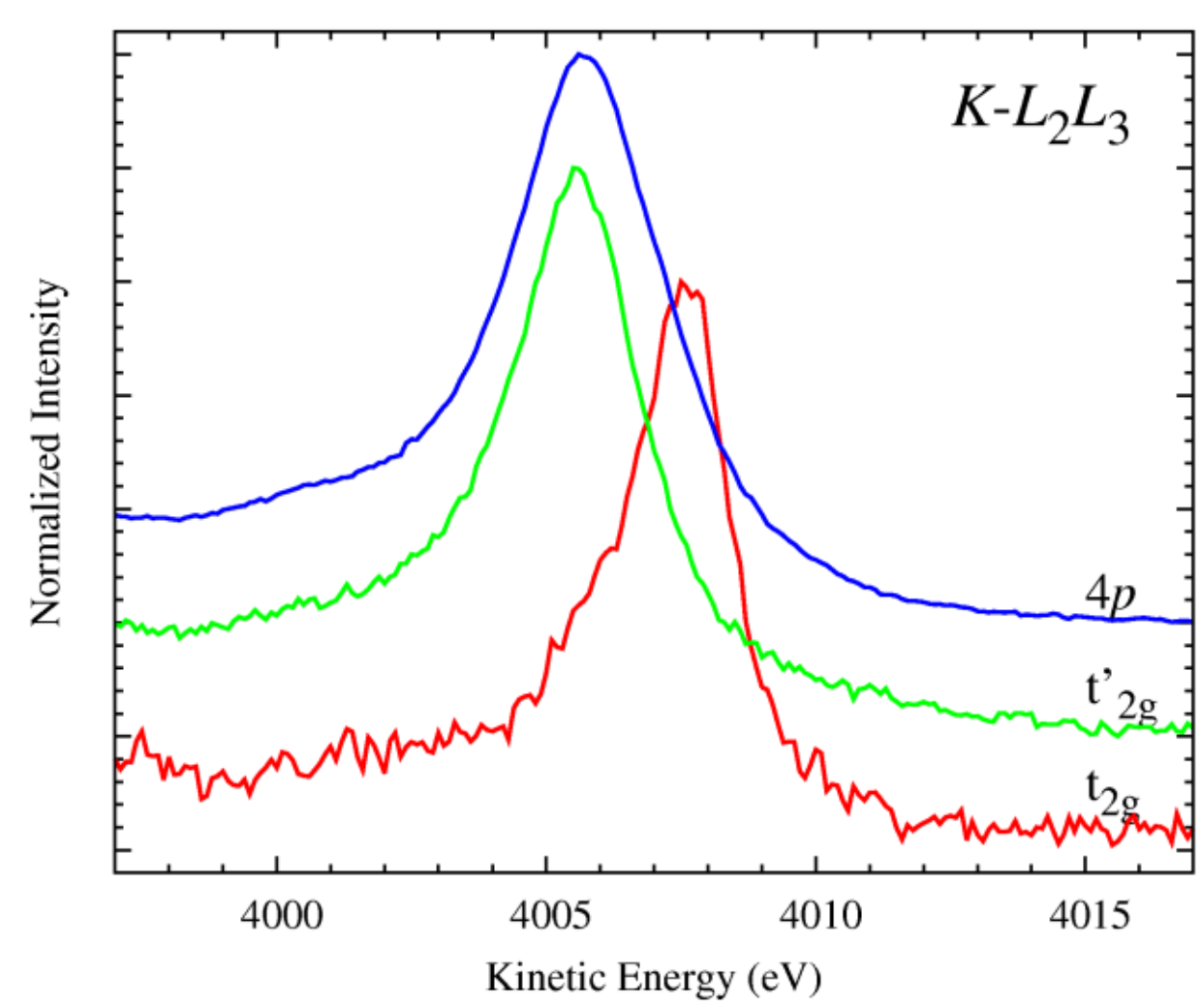

Fig.3. Resonant Ti *K*-$L_2L_3$ Auger spectra recorded with photon energy set to the Ti 1*s* → $t_{2g}$, 1*s* → $t'_{2g}$, and 1*s* → 4*p* transitions indicated in Figure 2. The data have been normalized to equal peak height.

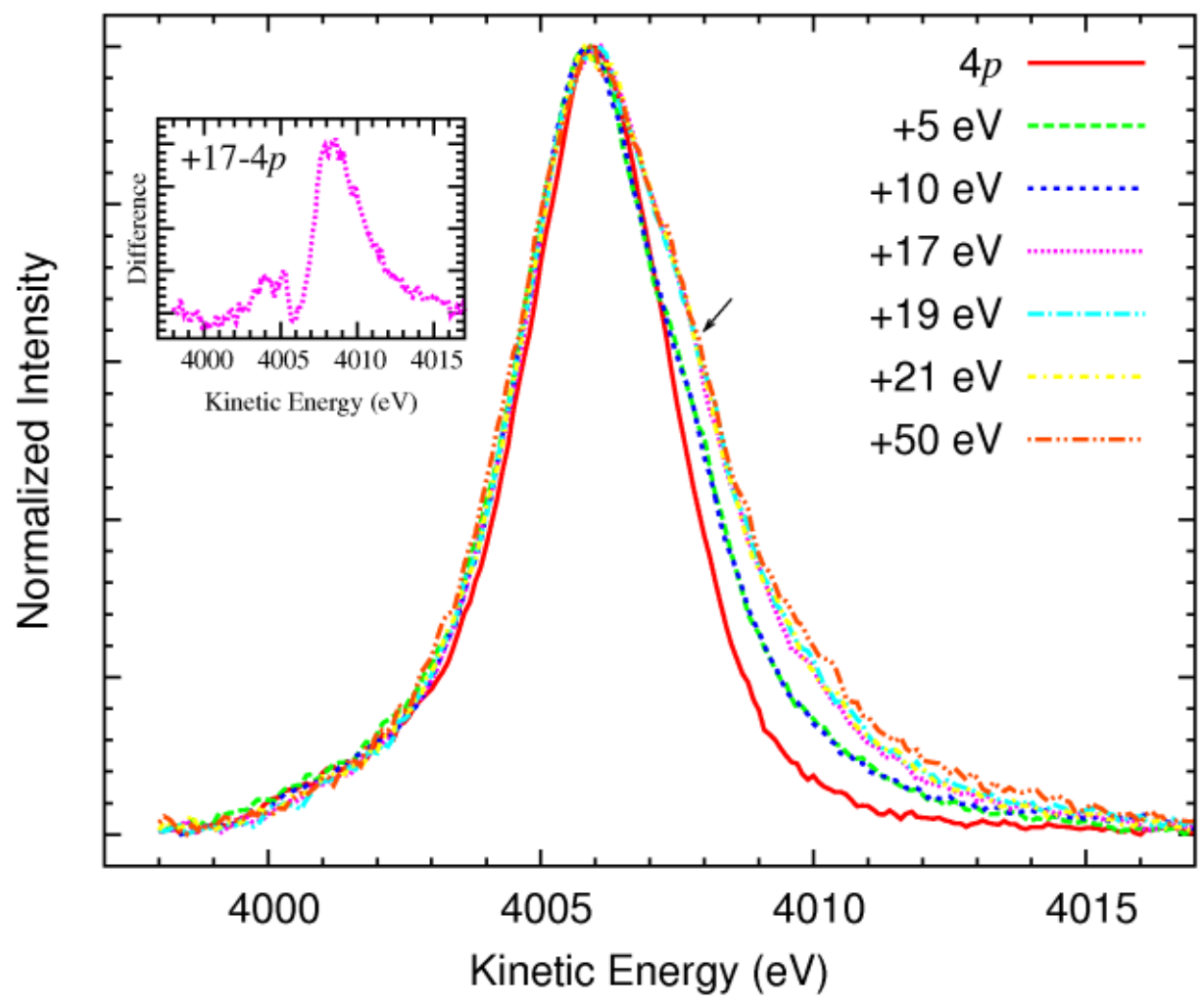

Fig.4. Ti $K$-$L_2L_3$ Auger spectra recorded with photon energy set to the Ti $1s \rightarrow 4p$ transition and with excess energy above threshold as indicated. Note the kink and the additional intensity on the high-kinetic energy side of the Auger peak that turns on discretely at 5 eV and then again at 17 eV excess photon energy. The inset shows the difference spectrum between the spectrum recorded at the $1s \rightarrow 4p$ transition and 17 eV above it. The data have been normalized to equal peak height and have had an integrated background removed.

Clearly, there is an additional feature at the high kinetic-energy side of the main Auger peak that turns on discretely with excess photon energies above the $1s \rightarrow 4p$ edge. From the difference spectrum shown in the inset, we find that the kinetic energy of this feature coincides with the kinetic energy of the Auger peak when the photon energy is set to either the $t_{2g}$ or $e_g$ local $1s \rightarrow 3d$ resonance; i.e., in the presence of the screening charge of the $t_{2g}$ or $e_g$ $3d$ spectator electron. We note that if this additional intensity were due to an intrinsic satellite or loss feature associated with the primary or diagram Auger decay, it would occur at a kinetic energy below the main line. Consequently, it must reflect the same well-screened initial states of the Auger decay but that turn on discretely at the excess photon energies above the $4p$ threshold that are equal to the binding energies of the two Ti photoelectron satellites; i.e., $E + \Delta E_1$ and $E + \Delta E_2$ where $E$ is the threshold for the core ionization and $\Delta E$ is the additional energy required for the "shake" [27]. Similar threshold phenomenon has been observed for $K$-$L_2L_3$ Auger decay of Ar gas [25] and Cu and Ni metals [28], but to our knowledge this is the first time that such a satellite has been observed in a solid on the high-kinetic (low-binding) energy side of the primary Auger peak clearly identifying it as a "shake-on" rather than a "shake-off" charge process.

Figure 5 illustrates the three possible transitions pertaining to the Auger decay: $K$-$L_2L_3$ Auger decay following direct $1s$ photoionization, $K$-$L_2L_3$ resonant Auger decay following promotion of the $1s$ electron to the unoccupied metal $3d$ orbitals, and $K$-$L_2L_3$ Auger decay following $1s$ photoionization and O $2p$ to Ti $3d$ ligand-to-metal charge transfer. Note the similarity (and hence the equality of final-state energies) of the initial states of the Auger decay in the latter two cases [29]; clearly, the high-energy structure observed in the Ti $K$-$L_2L_3$ Auger spectrum is due to the contribution from the well-screened initial states that are created by the $t_{2g}$ and $e_g$ ligand-to-metal charge transfer that occur for energies above the Ti $4p$ edge.

We will now discuss our data within the context of the available theoretical calculations. Ikemoto et al. [6] found satellite structure in the Ti $2p$ core ionization of Ti based transition-metal oxides and attributed it to O $2p \rightarrow$ Ti $4s$ shake-up. Kim and Winograd [7] inter-

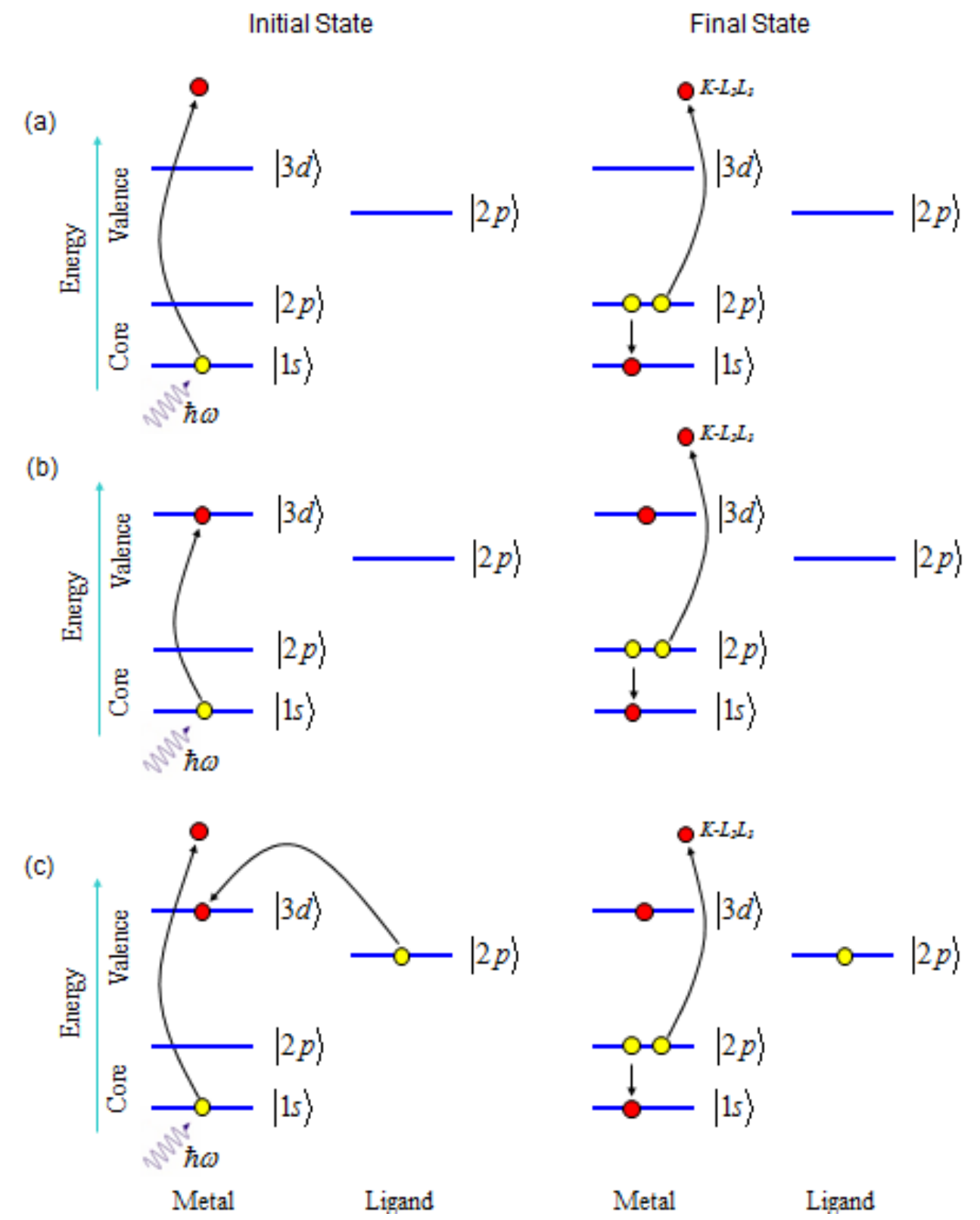

Fig. 5. Illustrations of initial and final states for: (a) Ti $K$-$L_2L_3$ Auger decay following $1s$ core photo-excitation that leaves the Ti ion doubly ionized; (b) Ti $K$-$L_2L_3$ Auger decay following resonant excitation of the $1s$ electron to the metal $3d$ orbitals; (c) Ti $K$-$L_2L_3$ Auger decay following $1s$ photo-excitation accompanied by charge transfer from the ligand O $2p$ to the metal Ti $3d$ orbitals. Note the similarity of the initial and final states for the Auger decays in (b) and (c) that leave the Ti ion singly ionized.

-preted similar data in terms of optical absorption spectra, and they assigned the observed satellites to monopole O $2p$ $e_g \rightarrow$ Ti $3d$ $e_g$, O $2p$ $a_{1g} \rightarrow$ Ti $4s$ $a_{1g}$, and O $2s$ $e_g \rightarrow$ Ti $3d$ $e_g$ transitions in order of increasing energy. These authors concluded that the first shake-up satellite occurring on the high binding-energy side of a transition-metal compound must be due to ligand $2p$ $e_g \rightarrow$ metal $3d$ $e_g$ monopole transitions because the probability of the first allowed shake-up, $2p$ $t_{2g} \rightarrow 3d$ $t_{2g}$, is too low to be observed. Sen et al. [8] disagreed with this hypothesis and argued that all such monopole transitions should be observed, and an anion-exciton model was later proposed by de Boer et al. [30] that challenged the conventional wisdom that the satellites are due to ligand-to-metal charge transfer. Since then, the 13 eV satellite has been reproduced by the full multiplet charge-transfer theory using a configuration-interaction wave function for a $TiO_6$ octahedral cluster and an Anderson-impurity Hamiltonian [8]. In these calculations, the on-site metal $d$-$d$ Coulomb repulsion energy $U$, the charge-transfer energy $\Delta$, and the ligand-metal $p$-$d$ hybridization energy $V$ are fitting parameters. This treatment was adopted by Bocquet et al. [10] and by Zimmermann et al. [11] who used a more complete configuration-interaction basis set but neglected both the core-hole $d$-electron multiplet effects and the crystal-field splitting of the metal $3d$ orbitals. These calculations consequently predict only the existence of the high-energy 13 eV structure.

Recently, Kas et al. [31] have applied an *ab initio* real-time cumulant approach for charge-transfer satellites in x-ray photoemission data. Their calculation reproduces the Ti $2p$ core lines and the high-energy 13 eV satellite. The 5 eV satellite is again either missed or obscured by the spin-orbit splitting of the Ti 2*p* level. A major advantage of this theoretical treatment over previous work, however, is that it is a real-space approach based on density-functional theory (DFT) in which the cumulant representation describes the transfer of spectral weight from the main quasi-particle or core peak to the satellite. Afforded by the calculation is the charge density of the 13 eV ligand-to-metal excitation that clearly identifies it as $e_g$ symmetry, bearing stunning resemblance to the $e_g$ set of molecular orbitals for this system [32].

From group theory and molecular-orbital considerations, electronic transitions with $e_g$ symmetry should naturally lie at lower-kinetic energy (higher-binding energy relative to the main line) than those with $t_{2g}$ symmetry. This is because the energy required to excite an electron from an occupied O 2*p* $e_g$ level to an unoccupied metal 3*d* $e_g$ level is significantly greater than the energy required to excite an electron from an occupied O 2*p* $t_{2g}$ level to an unoccupied metal 3*d* $t_{2g}$ level. Likewise, because the transition probability in the sudden approximation is given by the square of the overlap between the initial and final states (hence the monopole-selection rules), the $e_g$ transition will be more intense than the $t_{2g}$ transition, once again because the metal 3*d* $e_g$ orbitals point towards the ligand 2*p* $e_g$ orbitals. Consequently, it is clear that the smaller-intensity 5 eV satellite observed in our data is due to ligand O 2*p* $t_{2g} \rightarrow$ metal 3*d* $t_{2g}$ transitions and the larger-intensity 13 eV satellite is due to ligand O 2*p* $e_g \rightarrow$ metal 3*d* $e_g$ transitions. As the formal-charge state of the Ti ion in this material is $Ti^{4+}$, both transitions should be observable in high-resolution photoelectron spectra.

In conclusion, we have identified a previously un-resolved photoelectron satellite in the Ti 1*s* core-level spectrum of the transition-metal oxide $SrTiO_3$, and we have examined the photon-energy dependence of the Ti $K$-$L_2L_3$ Auger decay within the vicinity of the Ti *K* edge. Our data reveal a low binding-energy feature in the Auger spectrum that is concurrent in energy to the Auger peak measured at both the Ti 1*s* $\rightarrow t_{2g}$ and the Ti 1*s* $\rightarrow e_g$ quadrupole transitions. This feature turns on discretely with excess photon energies above the 4*p* threshold that are equivalent to the two satellite binding energies thereby uniquely identifying them as O 2*p* $t_{2g} \rightarrow$ Ti 3*d* $t_{2g}$ and O 2*p* $e_g \rightarrow$ Ti 3*d* $e_g$ monopole ligand-to-metal charge transfer. The presence of two distinct satellites is due to the crystal-field splitting of the Ti 3*d* orbitals, and this assignment is consistent with recent *ab initio* theoretical calculations that predict the energy, intensity, and charge density of the higher-energy $e_g$ excitation. This work therefore points to a new direction on how photoelectron-satellite structure and the threshold behavior of Auger spectra may be used to study chemical bonding and orbital occupation in the solid state.

This work was performed at the National Synchrotron Light Source which is supported by the U.S. Department of Energy. Additional support was provided by the National Institute of Standards and Technology. The authors thank Dr. Eric Shirley for useful discussions and sharing unpublished calculations at the Ti 1*s* near edge.